%% file: prob.tex
\documentclass[submission,copyright,creativecommons]{eptcs}
\providecommand{\event}{ACT 2019}

\usepackage{tikz-cd}
\usepackage{amssymb, amsmath}
\usepackage[scr]{rsfso}
\usepackage{sgame}
\usepackage{enumerate}
\usepackage{tikz}

\usepackage[thmmarks]{ntheorem}

\usepackage[all,2cell]{xy}

\usepackage{mathtools}

\usepackage{microtype}

\usepackage[T1]{fontenc}

\usepackage{stmaryrd}

\usepackage{hyperref}
\usepackage{cleveref}

\input{macros.tex}

\title{Compositional Game Theory with Mixed Strategies: Probabilistic Open Games Using a Distributive Law}
\def\titlerunning{Compositional Game Theory with Mixed Strategies}

\author{Neil Ghani
\institute{University of Strathclyde}
\email{neil.ghani@strath.ac.uk}
\and
Clemens Kupke
\institute{University of Strathclyde}
\email{clemens.kupke@strath.ac.uk}
\and
Alasdair Lambert
\institute{University of Strathclyde}
\email{alasdair.lambert@strath.ac.uk}
\and
Fredrik Nordvall Forsberg
\institute{University of Strathclyde}
\email{fredrik.nordvall-forsberg@strath.ac.uk}
}
\def\authorrunning{N. Ghani, C. Kupke, A. Lambert \& F. Nordvall Forsberg}
\def\copyrightholders{Ghani, Kupke, Lambert \& Nordvall Forsberg}

\date{July 1, 2019}

\begin{document}

\maketitle

\begin{abstract}
  \noindent We extend the open games framework for compositional game
  theory to encompass also mixed strategies, making essential use of
  the discrete probability distribution monad. We show that the
  resulting games form a symmetric monoidal category, which can be
  used to compose probabilistic games in parallel and sequentially. We
  also consider morphisms between games, and show that intuitive
  constructions give rise to functors and adjunctions between pure and
  probabilistic open games.
\end{abstract}

\section{Introduction}

The research project of open games aims to re-develop the foundations
of economic game theory using compositionality and category
theory~\cite{hedgesthesis}, building on e.g.\ the work of
Escard\'o and Oliva~\cite{sequentialGames}. A compositional framework was proposed by
Ghani~et al.~\cite{compositionalGames}, which included operators from which to
build games from smaller component games, and solution concepts such
as pure Nash equilibria. However many games that can be found even
early on in an undergraduate textbook on game theory (such as e.g.\
Leyton-Brown and Shoham~\cite{gameTheory}) fail to contain any equilibria, unless
probabilistic (so-called \emph{mixed}) strategies are allowed. In
contrast, already Nash~\cite{nash} proves that mixed strategy Nash
equilibria always exist for games with a finite number of players and
strategies.

In this work, we extend the framework of open games also to mixed
strategies. We use the discrete probability distribution monad on
$\Set$ (a baby version of the Giry monad~\cite{giry}) to incorporate
probability distributions. However simply moving to the Kleisli
category for this monad is not sufficient for our purposes, as that
would fail to capture mixed strategies without also demanding e.g.\
probabilistic play functions. Instead, we make sure to enrich the
framework of open games with measured use of the distribution monad in
the appropriate places. In particular, we construct a ``relational
Kleisli lifting'', a variant of the relational lifting for set
functors (cf.\ e.g.\ Kupke, Kurz and Venema~\cite{kupke2012lifting}), that turns
predicates with non-probabilistic parameters into predicates with
mixed parameters in a non-trivial way.

\section{Compositional Game Theory with Pure Strategies}

We briefly recall the definition of non-probabilistic, ``pure'' open
games as introduced by Hedges~\cite{hedgesthesis} for modelling economic
game theory with deterministic agents.

\begin{definition}
  Let $X$, $Y$, $R$ and $S$ be sets. A \emph{pure open game}
  $\G = (\Sigma_\G, P_\G, C_\G, \equilib_\G) : (X,S) \longrightarrow
  (Y,R)$ consists of:
  \begin{itemize}
  \item a set $\Sigma_\G$, called the set of \emph{strategy profiles}
    of $\G$,
  \item a function $P_\G : \Sigma_\G \times X \to Y$, called the
    \emph{play function} of $\G$,
  \item a function $C_\G : \Sigma_\G \times X \times R \to S$, called
    the \emph{coutility function} of $\G$, and
  \item a function
    $\equilib_\G : X \times (Y \to R) \to \Pow(\Sigma_\G)$, called the
    \emph{equilibrium function} of $\G$.
  \end{itemize}
\end{definition}
As these games are open,
they have an interface for interacting with other games. This consists of
a set $X$ representing the state/history of the game, a set $Y$ of
possible moves, a set $R$ of possible outcomes, and a set $S$ of
possible outcomes to feed back to the environment. Open games also have a strategy set $\Sigma_{\G}$ from which we wish to determine the optimal strategy.
The play function $P_{\G}$ produces a move based on the state and strategy. The coutility function $C_{\G}$ then determines which outcome is returned to the environment based on the state, strategy and outcome, and the equilibrium function $\equilib_{\G}$ determines which strategies are optimal given the state and utility function. See \Cref{ex:matching-pennies-prob} on the next page for an example. The game given there is probabilistic, but as we will see, most of the structure is shared between pure and probabilistic games.

The following fundamental theorem of pure open games allows parallel and sequential composition:

\begin{theorem}[{Ghani et al.~\cite{compositionalGames}}]
  The collection of pairs $(X, S)$ of sets $X$ and $S$, with pure open
  games $\G:(X,S) \longrightarrow (Y,R)$ as morphisms, forms a
  symmetric monoidal category $\Pure$. \hfill \QEDsymbol
\end{theorem}
To be precise, in order to satisfy the category axioms on the nose,
one needs to quotient by the equivalence relation induced by
isomorphism of strategies. We simplify presentation here and in what
follows by dealing with representatives directly.

\section{Probabilistic Open Games}

Our aim is to extend the framework of compositional game theory to
also encompass mixed strategies, i.e.\ games where players' strategies
are probability distributions over pure strategies. For a set $X$,
write $\D(X)$ for the set of discrete probability distributions on
$X$, i.e.\ $\D(X)$ is the collection of functions
$\omega : X \to [0,1]$ with $\sum_{x \in X}\omega(x) = 1$ whose
support $\supp(\omega) = \{x \in X\ |\ \omega(x) \neq 0\}$ is
finite. It is well known that $\D : \Set \to \Set$ is a monad (see
e.g.\ Jacobs~\cite{jacobsDoverview} for an overview of probability monads
in different categories), and we will make essential use of this
structure in the following. The unit of the monad $\eta : X \to \D X$
maps elements to point distributions, and the multiplication
$\mu :\D^2X \to \D X$ ``flattens'' a distribution of
distributions. Furthermore, $\D$ is a commutative strong monad,
meaning that there is a double strength natural transformation
$\laxell : \D{}A \times \D{}B \to \D(A \times B)$ given by forming the
independent joint distribution.
Algebras of $\D$ are convex sets, which we think of as
sets $R$ equipped with the operation of taking expected values
$\Expect : \D(R) \to R$. We do not expect all sets involved in a game
to support this operation --- e.g.\ the set of moves is typically
discrete --- but we do expect (and need) the sets of possible outcomes
for the games and its environment to do so.

\begin{definition}
  Let $X$, $Y$ be sets, and $R$, $S$ be $\D$-algebras. A
  \emph{probabilistic open game}
  $\G = (\Sigma_\G, P_\G, C_\G, \equilib_\G) : (X,S) \longrightarrow
  (Y,R)$ consists of:
  \begin{itemize}
  \item a set $\Sigma_\G$, called the set of \emph{strategy profiles}
    of $\G$,
  \item a function $P_\G : \Sigma_\G \times X \to Y$, called the
    \emph{play function} of $\G$,
  \item a function $C_\G : \Sigma_\G \times X \times R \to S$, called
    the \emph{coutility function} of $\G$, and
  \item a function
    $\equilib_\G : X \times (Y \to R) \to \Pow(\D(\Sigma_\G))$, called
    the \emph{equilibrium function} of $\G$.
  \end{itemize}
\end{definition}

In other words, a probabilistic open game consists of the same data as
a pure open game, except that the equilibrium function records which
\emph{mixed} strategies are ``optimal'', instead of just being
concerned with pure strategies.  Overall, this matches how we usually
think of games with mixed strategies: the moves and outcomes of the
game stays the same, only the strategies can be probabilistic. The
$\D$-algebra structure of $R$ and $S$ is not needed for this basic
definition, but will be used to compose games.

\begin{figure}[tbh]
\begin{center}
\begin{game}{2}{2}
      & $H$     & $T$\\
$H$   & $-1,1$   & $1,-1$\\
$T$   & $1,-1$   & $-1,1$
\end{game}
\end{center}
\caption[]{Utility $k$ of the Matching Pennies game.}
\label{fig:oneround_util}
\end{figure}

\begin{example}
  \label{ex:matching-pennies-prob}
  The Matching Pennies game involves two players trying to win pennies
  from each other. Each player puts forward one side of a penny, heads
  or tails. If the faces match then the first player wins the second
  player's penny, and if they do not match, the second player instead
  wins the first player's penny. This is summarised in
  \Cref{fig:oneround_util}. We can represent Matching Pennies as a
  state-free open game
  \[
     \mathcal{MP}:(\One, \R \times \R) \longrightarrow (\{H, T\} \times \{H,T\}, \R \times \R)
  \]
  with utility and coutility taken from $\R \times \R$, and moves
  $Y \times Y$ where $Y = \{H, T\}$ --- each player either plays heads
  or tails. A pure strategy is simply a move (i.e.\ the strategy set
  for the game is $\Sigma_{\mathcal{MP}} = Y \times Y$), hence both
  the play and coutility functions $P_{\mathcal{MP}}$ and
  $C_{\mathcal{MP}}$ are particularly simple, given by
  $P_{\mathcal{MP}}(c) = c$ and $C_{\mathcal{MP}} (c,r) = r$
  respectively.  The equilibrium
  $\equilib_{\mathcal{MP}}: {(Y \times Y \to \R \times \R)} \to \Pow
  (\D (\Sigma_{\mathcal{MP}}))$ is defined by
  $\phi \in \equilib_{\mathcal{MP}}\, k$ if and only if
  \begin{align*}
     & \phi_1 \in  \argmax\limits_{\phi_{1}' \in \D  Y}(\Expect [\D(\lambda y\phantom{'}.\Expect [ \D (\pi_{1} k(y\phantom{'}, -)) \phi_2])\phi_{1}']) \\
      \text{ and } \; &\phi_2  \in \argmax\limits_{\phi_{2}' \in \D  Y}(\Expect [ \D(\lambda y' .\Expect [ \D(\pi_{2} k(-, y'))\phi_1])\phi_{2}'])
  \end{align*}
  where $\phi_i = \D(\pi_i)\phi$ are the \emph{marginals} of
  $\phi$. We see that both players are trying to maximise their
  expected payoff, assuming their opponent probabilistically plays
  according to their fixed strategy.
\end{example}

\section{Probabilistic Open Games Form a Symmetric Monoidal Category}

Just like pure open games, probabilistic open games support a wide
range of operations: they can be composed in parallel, composed
sequentially, conditioned, iterated, and much more. Here we focus on
parallel and sequential composition, and prove that these operations
make the collection of pairs of sets with probabilistic open games as
morphisms a symmetric monoidal category.

\subsection{Parallel composition of probabilistic open games}

The parallel composition represents two games played simultaneously.
Its definition makes crucial use of the fact that the category of $\D$-algebras has all limits, since it employs products of $\D$-algebras $R \times R'$ and $S \times S'$.

\begin{definition}
  Let $\G:(X,S) \longrightarrow (Y,R)$ and
  $\G':(X', S') \longrightarrow (Y',R')$ be probabilistic open
  games. We define the \emph{parallel composition} probabilistic open
  game
  $\G \otimes \G' : (X \times X', {S \times S'}) \longrightarrow (Y
  \times Y',R \times R')$ as follows:
  \begin{itemize}
  \item the strategy set is
    $\Sigma_{\G \otimes \G'} = \Sigma_{\G} \times \Sigma_{\G'}$;
  \item the play function is defined by
    $P_{\G \otimes \G'}((\sigma, \sigma'), (x, x')) = (P_{\G}(\sigma,
    x), P_{\G'}(\sigma', x'))$;
  \item the coutility function is defined by
    $C_{\G \otimes \G'}((\sigma, \sigma'), (x, x'), (r,r')) =
    (C_{\G}(\sigma, x, r), C_{\G'}(\sigma', x', r'))$;
  \item the equilibrium function
    $\equilib_{\G \otimes \G'}: (X\times X') \times (Y \times Y' \to R \times
    R') \to \Pow(\D(\Sigma_{\G} \times \Sigma_{\G'}))$ is defined by
    \begin{align*}
      \equilib_{\G \otimes \G'} \; (x_1, x_2) \; k = \{\ \laxell(\phi_1, \phi_2)\ |\ \ & \phi_1 \in \equilib_{\G} \, x_1\; \Expect[\D(\pi_1 \circ k) \circ \laxell(\eta- , \D(P_{\G'}(-,x_2))\phi_2)] \wedge {}
      \\ & \phi_{2} \in \equilib_{\G'} \, x_2 \; \Expect[\D(\pi_2 \circ k) \circ \laxell(\D({P}_{\G}(-,x_1)\phi_1), \eta-)]\ \} \ \ \ \ \ \ \ \ \ \  \blacklozenge
    \end{align*}
    \end{itemize}
\end{definition}

The definition of the strategy set, play function and coutility
function coincides with the definition of parallel composition for
pure open games, as expected. The equilibrium function of the parallel
game is more complicated because of the probabilities involved --- note
that this makes essential use of the $\D$-algebra structure on $R$.
Basically, each player is trying to find an equilibrium for the
utility function which computes the expected utility for the original
utility function $k : Y \times Y' \to R \times R'$, assuming the other
player plays \emph{probabilistically} using their fixed strategy.
Note that even though $\equilib_{\G \otimes \G'} \; (x_1, x_2) \; k$ is a
predicate on $\D(\Sigma_{\G} \times \Sigma_{\G'})$, and not on
$\D(\Sigma_\G) \times \D(\Sigma_{\G'})$, only strategies that arise
independently from strategies in $\Sigma_{G}$ and $\Sigma_{\G'}$ are
in the equilibrium. Game-theoretically, this makes sense, as the
players are not expected to cooperate, and mathematically, this is
crucial for parallel composition to be associative.

\begin{example}
  In \Cref{ex:matching-pennies-prob} we showed that the Matching
  Pennies game can be represented as a open game. We now show that we
  can build this game as the parallel composition of two identical
  component ``player'' games
  $\mathcal{MP}_{i}:(\One,\R) \longrightarrow (\{H,T\}, \R)$.
  Strategies are moves $\Sigma_{\mathcal{MP}_i} = Y = \{H,T\}$ and the
  play function is given as the identity on strategies. The coutility
  is given as the second projection returning the utility. Finally the
  equilibrium function $\equilib_{\mathcal{MP}_i} : (Y \to \R) \to
  \Pow(\D(\Sigma_{\mathcal{MP}_i}))$ is given by
  \[
    \phi  \in \equilib_{\mathcal{MP}_i} k\, \text{ if } \, \phi \in \argmax(\Expect [ \D(k)(-)])
  \]
  i.e.\ a mixed strategy is optimal if it maximises the expected
  payoff.  The parallel composition of $\mathcal{MP}_{1}$ and
  $\mathcal{MP}_{2}$ produces the Matching Pennies game described in
  \Cref{ex:matching-pennies-prob}
  \[
    \mathcal{MP}_{1}\otimes \mathcal{MP}_{2} \cong \mathcal{MP} \enspace .
  \]
  The equilibrium function for the composed game states that
  $\phi \in \equilib_{\mathcal{MP}_{1}\otimes \mathcal{MP}_{2}}\, k$
  if 
  \begin{align*}
    & \phi_{1} \in \equilib_{\mathcal{MP}_{1}}(\lambda y\phantom{'}. \Expect [ \D (\pi_{1} \circ k)\laxell (\eta(y)\phantom{'}, \phi_{2}) ])\\
    \text{ and } & \phi_{2}\in \equilib_{\mathcal{MP}_{2}}(\lambda y'. \Expect [ \D (\pi_{2} \circ k)\laxell (\phi_{1}, \eta(y')) ])
  \end{align*}
  where $\phi_i = \D(\pi_i)\phi$ are the marginals of $\phi$.

  To show that our definition gives the expected results from economic
  game theory, we now solve this game, i.e.\ we compute a more
  concrete description of $\equilib_{\mathcal{MP}}\;k$ for the utility
  function from \Cref{fig:oneround_util}. As Matching Pennies is a
  symmetric game we focus on the first player's equilibrium. Expanding
  the definition of $\equilib_{\mathcal{MP}_{1}}$, the condition says
  \[
    \phi_{1}  \in \argmax\limits_{\phi_{1}' \in \D \Sigma}(\Expect [\D(\lambda y.\Expect [ \D (\pi_{1} \circ k)\laxell (\eta(y),  \phi_{2})])\phi_{1}'])
  \]

  The vigilant reader might have noticed that the equilibrium
  condition here is not syntactically the same as the one given in
  \Cref{ex:matching-pennies-prob}, but because of the point
  distributions $\eta(y)$ involved, it is not hard to see that the
  expressions are equal. Reducing the terms down and instantiating the
  utility function from \Cref{fig:oneround_util}, we reach
  \[
    \phi_{1}  \in \argmax\limits_{\phi_{1}' \in D \Sigma}(\sum\limits_{r \in R}r \sum\limits_{\{y \in Y | \phi_{2}(y) - \phi_{2}(\bar{y}) =r\}} \phi_{1}'(y))
  \]
  As there are only two pure strategies, we can consider both
  possibilities for $\phi_2$ in terms of $\phi_2(H)$ only:
  \begin{align*}
    \phi_2(H) - \phi_2(T) & = \phi_{2}(H) - (1 - \phi_{2}(H))&    \phi_{2}(T) - \phi_{2}(H) & = (1 - \phi_{2}(H)) - \phi_{2}(H)\\
    & = 2\phi_{2}(H) - 1&    & = 1 - 2\phi_{2}(H)
  \end{align*}
  Rearranging and substituting into the formula, we arrive at the
  condition
  \[
    \phi_{1} \in \argmax\limits_{\phi'}((2\phi_2(H) -1)(2\phi'(H)-1))
  \]
  and since the game is symmetric we similarly obtain for the second player
  \[
    \phi_{2} \in \argmax\limits_{\phi''}((2\phi_1(H) -1)(1 -2\phi''(H)))
  \]
  leaving three cases to consider:
  \begin{align*}
    \text{if  } & \phi_2(H)=1/2 \, \Rightarrow \, \phi_1(H)\in [0,1]  & \phi_1(H)=1/2 \, \Rightarrow \, \phi_2(H)\in[0,1]\\
   & \phi_2(H)< 1/2 \, \Rightarrow \, \phi_1(H) = 0 \quad \, & \phi_1(H) < 1/2\, \Rightarrow \, \phi_2(H)=1\\
   & \phi_2(H)> 1/2 \, \Rightarrow \, \phi_1(H) = 1 \quad \, & \phi_1(H)> 1/2 \, \Rightarrow \, \phi_2(H)=0
  \end{align*}
  The only point of stability 
  lies at $\phi_1(H) = \phi_2(H) = 1/2$, since if one player deviates
  from this strategy the other will return the favour. Hence the only
  equilibrium is for both players to play both strategies with $50\%$
  probability, indeed the standard solution.
\end{example}

In order to prove associativity of parallel composition, we use a
``determinisation'' construction that turns probabilistic games into
pure games, reminiscent of the abstract categorical formulation of
automata determinisation presented e.g. in Silva et al.~\cite{siboboru:gene13}.
This way, we can reuse part of the proof that parallel composition is
associative for pure games~\cite{compositionalGames}.

\begin{definition}
  \label{def:determ-game}
  Given a probabilistic game $\G : (X,S) \longrightarrow (Y,R)$ with
  strategy set $\Sigma$, we define its \emph{determinisation} pure
  game $\hatgame{\G}: (X, S)\longrightarrow (\D Y,\D R)$ with strategy
  set $\D\Sigma$ and
  \begin{itemize}
  \item play function
    $P_{\hatgame{\G}}(\phi, x) = \D(P_{\G}(-,x))\phi$;
  \item coutility function
    $\, C_{\hatgame{\G}}(\phi, x, \psi) =
    \Expect[\D(C_{\G}(-,x,-))\laxell(\phi, \psi)]$; and
  \item equilibrium function
    $\phi \in \equilib_{\hatgame{\G}}\, x \; k$ if and only if
    $\phi \in \equilib_{\G}\, x \; (\Expect \circ k \circ \eta)$.
  \end{itemize}
\end{definition}

Using the naturality of $\eta$, and that $\Expect : \D(R) \to R$ is a
$\D$-algebra, it is easy to see the following way to go between the
equilibria of $\G$ and $\hatgame{G}$:

\begin{lemma}
  \label{thm:determine-Dk}
  Let $k : Y \to R$. Then
  $\phi \in \equilib_{\hatgame{\G}}\, x \; \D(k)$ if and only if
  $\phi \in \equilib_{\G}\, x \, k$. \hfill \QEDsymbol
\end{lemma}

In general, it is not the case that the determinisation of a parallel
composition is a parallel composition of determinisations --- for
instance, the type of moves do not even match up, since in general
$\D(Y \times Y') \not\cong \D{}Y \times \D{}Y'$. To obtain even a lax
monoidal map
$\hatgame{\G} \otimes \hatgame{\G '} \to \hatgame{\G \otimes \G '}$,
we need to restrict to utility functions that respect the $\D$-algebra
structure, which for instance Kleisli extensions do. This is
formulated in the following lemma.

\begin{lemma}
  \label{thm:Determine-otimes}
  Let $\G: (X,S) \longrightarrow (Y,R)$ and
  $\G':(X', S') \longrightarrow (Y',R')$ be probabilistic open games.
  For all $\phi \in \D\Sigma_{\G} \times \D\Sigma_{\G'}$, $x \in X \times X'$, and
  $k : Y \times Y' \to \D(R \times R')$, we have
  \[
    \laxell(\phi) \in \equilib_{\hatgame{\G \otimes \G'}} \, x \; \KleisliLift{k} \text{ iff }
    \phi \in \equilib_{\hatgame{\G} \otimes \hatgame{\G '}}\, x\, (\langle \D(\pi_1), \D(\pi_2)\rangle \circ \KleisliLift{k} \circ \laxell)
  \]
  where
  $\KleisliLift{k} = \mu \circ \D(k) : \D(Y \times Y') \to \D(R \times
  R')$ is the Kleisli extension of $k$. \hfill \QEDsymbol
\end{lemma}

We use this lemma to prove the
associativity of parallel composition of probabilistic games using the
corresponding associativity for pure games.

\begin{theorem}
  \label{thm:assoc-parallel}
  Let $\G:(X,S) \longrightarrow (Y,R)$,
  $\G':(X', S') \longrightarrow (Y',R')$ and
  $\G'' : (X'', S'') \longrightarrow (Y'',R'')$ be probabilistic open
  games. We have
  $\G\otimes(\G' \otimes \G'') = (\G\otimes\G') \otimes \G''$, up to
  canonical isomorphisms
  $A \times (A' \times A'') \cong (A \times A') \times A''$ of the
  underlying sets involved. \hfill \QEDsymbol
\end{theorem}

\subsection{Sequential composition of probabilistic open games}

Another fundamental operation to modularly build games is sequential
composition. Intuitively, in the sequential composition $\seq{\G}{\HH}$ of
open games $\G$ and $\HH$, we first play $\G$, followed by $\HH$. This
means the moves of $\G$ are the states of $\HH$, and pure strategies of
$\seq{\G}{\HH}$ are pairs of pure strategies for $\G$ and $\HH$. A mixed
strategy $\phi$ of the composed game $\seq{G}{H}$ is an equilibrium if
the marginal distributions are equilibria in $\G$ (relative to the
payoff function for $\G$ that we obtain by feeding $\HH$'s coutility
back) and $\HH$ (relative to the given payoff function of $\seq{\G}{\HH}$),
respectively. In order to state the latter, we first need to define a
``Kleisli predicate lifting'' of
$\equilib_{\HH}(-,k) : Y \to \Pow(\D(\Sigma_{\HH}))$, since we only get a
mixed state in $\D(Y)$ as a result of probabilistically playing the
first game using the first mixed strategy.

\begin{definition}
  \label{def:liftpred}
  Let $R : X \to \Pow(\D(Y))$. We define
  $\liftpred{R} : \D(X) \to \Pow(\D(Y))$ by
  $\liftpred{R} = \Pow(\mu_{Y}) \circ \lambda_{\D(Y)} \circ \D(R)$, where
  $\lambda : \D\Pow \to \Pow\D$ is the transformation given by
  \[
    \lambda_{X}(\alpha) = \{ \phi \in \D{}X|\ \big(\exists \rho \in \D(\in \subseteq X \times \Pow{}X)\big)\big(\D(\pi_1)\rho = \phi \text{ and } \D(\pi_2)\rho = \alpha\big) \} \enspace .
  \]
\end{definition}

Concretely, for $\alpha = \sum_{i}p_ix_i \in \D(X)$, we have
\[
  \liftpred{R}(\alpha) = \{ \mu(\sum_{i}\sum_{j}q_{i,j}\psi_{i,j}) \ |\ \sum_{j}q_{i,j} = p_i \text{ and } \psi_{i,j} \in R(x_i) \}
\]
where $\sum_{i}p_i\phi_i$ is the distribution on $Y$ assigning
probability $\sum_{i}p_i\phi_i(y)$ to $y \in Y$.
By the abstract definition, we immediately have that
$\liftpred{R \circ f} = \liftpred{R} \circ \D(f)$ since
$\D$ is a functor.
We now use this lifting to define the sequential composition of two
probabilistic games.

\begin{definition}
  Let $\G:(X,S) \longrightarrow (Y,R)$ and
  $\HH : (Y, R) \longrightarrow (Z,T)$ be probabilistic open games. We
  define the \emph{sequential composition} probabilistic open game
  $\seq{\G}{\HH} :(X,S) \longrightarrow (Z,T)$ as follows:
  \begin{itemize}
  \item the strategy set is
    $\Sigma_{\seq{\G}{\HH}} = \Sigma_\G \times \Sigma_\HH$;
  \item the play function is defined by
    $P_{\seq{\G}{\HH}}((\sigma_1,\sigma_2),x) =
    P_\HH(\sigma_2,P_\G(\sigma_1,x))$;
  \item the coutility function is defined by
    $C_{\seq{\G}{\HH}}((\sigma_1,\sigma_2),x,t) =
    C_\G(\sigma_1,x,C_\HH(\sigma_2,P_\G(\sigma_1,x),t))$;
  \item the equilibrium function
    $\equilib_{\seq{\G}{\HH}} : X \times (Z \to T) \to
    \Pow(\D(\Sigma_{\G}\times \Sigma_{\HH})) $ is defined by
  \begin{align*}
    \equilib_{\seq{\G}{\HH}}\, x \, k = \{\ \laxell(\phi_1, \phi_2)\ |\ \ & \phi_{1}\in  \equilib_{\G}\; x \; (\lambda y.\, \Expect [ \D(\lambda \sigma.\, C_{\HH}(\sigma, y, k(P_{\HH}(\sigma, y))))\phi_2]) \wedge {} \\
                                                                   & \phi_2 \in  \liftpred{\equilib_{\HH}(-,k)}\, (\D(P_{\G}(-, x))\phi_{1}) \} \qquad \qquad \qquad \qquad \ \ \ \ \ \ \ \ \ \ \ \ \ \ \  \blacklozenge
  \end{align*}
\end{itemize}
\end{definition}

To see that this definition is meaningful game-theoretically, we model
the well-known Market Entry game (Stackelberg \cite{stackelbergTrans}) using our
framework.

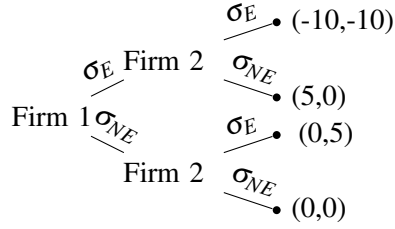
\begin{figure}[tb]
\begin{center}
\tikzstyle{level 1}=[level distance=3cm, sibling distance=3cm]
\tikzstyle{level 2}=[level distance=3cm, sibling distance=2cm]
\tikzstyle{firm} = [text width=4em, text centered]
\tikzstyle{end} = [circle, minimum width=3pt,fill, inner sep=0pt]
  \begin{tikzpicture}[grow=right, sloped,scale = 0.50]
\node[firm] {Firm 1}
    child {
        node[firm] {Firm 2}
            child {
                node[end, label=right:
                    {(0,0)}] {}
                edge from parent
                node[above] {$\de$}
                node[below]  {}
            }
            child {
                node[end, label=right:
                    { (0,5)}] {}
                edge from parent
                node[above] {$\e$}
                node[below]  {}
            }
            edge from parent
            node[above] {$\de$}
            node[below]  {}
    }
    child {
        node[firm] {Firm 2}
        child {
                node[end, label=right:
                    {(5,0)}] {}
                edge from parent
                node[above] {$\de$}
                node[below]  {}
            }
            child {
                node[end, label=right:
                    {(-10,-10)}] {}
                edge from parent
                node[above] {$\e$}
                node[below]  {}
            }
        edge from parent
          node[above] {$\e$}
           node[below]  {}
    };
\end{tikzpicture}
\caption{Market Entry game: firms 1 and 2 decide whether to enter ($\e$) or not enter ($\de$) the market.}
\label{fig:market-entry}
\end{center}
\end{figure}

\begin{example}
\label{ex:ME}
The market entry game models two competing firms wishing to enter a
new market. If they both enter, the competition between them would be
expensive. The situation is depicted in \Cref{fig:market-entry}.  Firm
1 enters first, firm 2 then observes the move made and responds. If
one firm enters alone they will reap the rewards, but if both enter
they will both suffer. Of course if neither enters then nothing
happens. We expect the only subgame perfect equilibrium to be where
the first firm enters, and the second firm reverses the first firm's
decision.

We model this as a sequential composition $\seq{\G_1}{\G_2}$ of two
probabilistic open games.  The first game
$\G_1:(\One,\R\times \R) \longrightarrow ( \{\e, \de\}, \R \times \R) $
has strategy set $\Sigma_1 = Y = \{\e,\de\}$ the set of moves, and the
obvious play and coutility functions.  Its equilibria are
\begin{equation*}
\equilib_1 \, (k : Y \to \R \times \R) = \argmax\limits_{\phi  \in \D \Sigma_1}\{ \Expect[\D(\pi_1 \circ k)(\phi)]\}
\end{equation*}
The second game
$\G_2:(\{\e, \de\}, \R \times \R) \longrightarrow (Y \times Y, \R
\times \R)$ arises as a ``subgame conditioned''
game~\cite[Def.~5]{iteratedGames} in order to allow the strategies
$\Sigma_2 = Y \to \Sigma_1$ to depend on the move made in $\G_1$.  The
play and coutility functions are given by $P_{2}(g, x) = (x, g(x))$
and $C_{2}(g, x, r) = r$. The equilibrium function insists on subgame
perfect strategies:
\begin{eqnarray*}
  \psi \in \equilib_2 \, y \, (k:Y\times Y \to \R \times \R) & \text{iff} &  \big(\forall y'\in Y\big) \ \D(\eval(-,y'))\psi \in \argmax\limits_{\psi'  \in \D \Sigma_1} \{\Expect [\D(\pi_2 \circ k(y', -))(\psi')] \}
\end{eqnarray*}
where $\eval : (A \to B) \times A \to B$ is function evaluation.

The sequential composition
$\seq{\G_1}{\G_2} :(\One,\R\times \R)\longrightarrow (Y \times Y, \R
\times \R)$ has as strategies pairs of strategies from each round
$\Sigma_{\seq{\G_1}{\G_2}} = \Sigma_1 \times \Sigma_2$.  For mixed
strategies $\phi \in \D \Sigma_1$ and $\psi\in \D\Sigma_2$, we have
$\laxell(\phi, \psi) \in \equilib_{\seq{\G_1}{\G_2}}\, (k:Y\times Y
\to \R\times \R)$ if and only if
\begin{align*}
  \phi &\in \equilib_{1}\;  (\lambda y.\, \Expect [ \D(\lambda f.\, k(y, f(y)))\psi])
  \mbox{ and } \\
  \psi &\in \liftpred{\equilib_{2}(-,k)}\phi = \D(\equilib_{2}(\e,k)) = \D(\equilib_{2}(\de,k))
\end{align*}
where the second condition has been simplified since
$\equilib_{2}(y,k)$ is independent of $y$.  For the utility function
$k$ from \Cref{fig:market-entry}, we further see that in fact
$\equilib_{2}(\e,k) = \equilib_{2}(\de,k) = \{ 1 \cdot \swap \}$ where
$\swap : Y \to Y$ is the function which swaps $\e$ and $\de$. Hence
for $\laxell(\phi, \psi) \in \equilib_{\seq{\G_1}{\G_2}}\, k$ we must
have $\psi = 1 \cdot \swap$ which in turns forces $\phi = 1 \cdot \e$
--- the expected (non-mixed) subgame perfect equilibria. Reflecting on
the argument, we see that our reasoning is an instance of backward
induction (see e.g.\ Leyton-Brown and Shoham ~\cite[\S4.4]{gameTheory}). It is interesting, but currently not clear to us, what the
$\liftpred{-}$ construction does in general when the second game has
not been conditioned to respond to the moves of the first game.
\end{example}

It is important to note that the distributive law
$\lambda : \D\Pow \to \Pow\D$ used in \Cref{def:liftpred} is
\emph{not} a distributive law between monads, because no such law
exists (Zwart and Marsden~\cite{zwart2018noLaw}). In particular, $\lambda$ does not
preserve the monad structure of $\D$, for instance
${\lambda_X \circ \eta_{\Pow{}X}} \neq \eta^\Pow_{\D{}X}$. It is
however a distributive law between functors (even of a functor over
the monad $\Pow$, and also over $\Pow^{\mathsf{op}}$, although we do not make use
of this fact), which will be important for us for proving
associativity of sequential composition.

\begin{fact}
  The transformation $\lambda : \D\Pow \to \Pow\D$ is is a
  distributive law between functors, i.e.\ it is natural. \hfill \QEDsymbol
\end{fact}
For a proof see Kupke, Kurz and Venema~\cite{kupke2012lifting}. Using the naturality of
$\lambda$, we can show that if $R : X \to \Pow\D{}Y$ and
$f : \D{}Y \to \D{}Y'$, then
$\Pow(f) \circ \liftpred{R} = \liftpred{\Pow(f) \circ R}$. This, with
$f$ being a marginal $\D(\pi)$, is one of the key steps to prove
associativity of
composition.

\begin{theorem}
  \label{thm:assoc-seq}
  Let $\G:(X,S) \longrightarrow (X',S')$,
  $\G':(X', S') \longrightarrow (X'',S'')$ and
  $\G'' : (X'', S'') \longrightarrow (Y,R)$ be probabilistic open
  games. We have
  $\seq{\G}{(\seq{\G'}{\G''})} = \seq{(\seq{\G}{G'})}{\G''}$, up to
  the canonical isomorphism
  $\Sigma_{\G} \times (\Sigma_{\G'} \times \Sigma_{\G''}) \cong (\Sigma_{\G} \times \Sigma_{\G'}) \times \Sigma_{\G''}$ of strategy sets. \hfill \QEDsymbol
\end{theorem}

\subsection{A symmetric monoidal category}

We have now assembled most of pieces needed to show that probabilistic
open games are the morphisms of a monoidal category: missing are unit
and identity games.

For each set $X$ and $\D$-algebra $S$, we define a probabilistic open
game $\Id_{(X,S)} : (X, S) \longrightarrow (X, S)$ with strategy set
$\Sigma_{\Id_{(X,S)}} = \One$, play function
$P_{\Id_{(X,S)}}(\sigma, x) = x$, coutility function
$C_{\Id_{(X,S)}}(\sigma, x,s) = s$, and equilibrium function
$\equilib_{\Id_{(X,S)}}\,x\,k = \One$, i.e.\ every (trivial) strategy
is an equilibrium.

\begin{lemma}
  \label{thm:prob-category}
  There is a category $\Prob$, where objects are pairs $(X, S)$ of a
  set $X$ and a $\D$-algebra $S$, and morphisms are probabilistic open
  games. Composition is given by sequential composition
  $\G \circ \HH = \seq{\HH}{\G}$, and the identity on $(X,S)$ is
  $\Id_{(X,S)}$. \hfill \QEDsymbol
\end{lemma}

Similarly, we define a trivial game
$\I : (\One, \One) \longrightarrow (\One, \One)$ with strategy set
$\Sigma_\I = \One$, the only possible play and coutility functions,
and equilibrium function $\equilib_{\I}\,x\,k = \One$, i.e.\ every
strategy is again an equilibrium.

\begin{lemma}
  \label{thm:bifunctor}
  The game $\I$ is the unit for parallel composition.
  Furthermore, the operation which maps $(X,S)$ and $(X', S')$ to
  $(X \times X',S \times S')$, and games $\G$ and $\G'$ to
  $\G \otimes \G'$, defines a bifunctor
  $\otimes : \Prob \times \Prob \to \Prob$. \hfill \QEDsymbol
\end{lemma}

Observing that $\Prob$ also has a symmetry (inherited from
$\Set \times (\D\text{-}\mathsf{Alg})^{\mathsf{op}}$), we have now
proved the following:

\begin{theorem}
  \label{thm:fundamental}
  The collection of pairs $(X, S)$ of a set $X$ and a $\D$-algebra
  $S$, with probabilistic open games $\G:(X,S) \longrightarrow (Y,R)$
  as morphisms, forms a symmetric monoidal category $\Prob$. \hfill
  \QEDsymbol
\end{theorem}

\section{Relating pure and probabilistic games}

We now construct a category where probabilistic open
games are the objects, by defining a notion of morphism between
games. In light of \Cref{thm:fundamental}, these morphisms are 2-cells
in a monoidal double category of games (cf.\
Hedges~\cite{julesMorphisms}). The construction works similarly for pure
games.  We then use the resulting categorical structure to relate pure
and probabilistic games in the form of an adjunction between the
categories.

As noticed by Ghani et al.~\cite{compositionalGames}, the definition of pure
open games can be given more compactly by employing the language of
lenses~\cite{lenses}.  A lens $(v, u) : (X,S) \to (Y,R)$ between pairs
of sets $(X, S)$ and $(Y, R)$ is given by a two functions
$v : X \to Y$ (``view'') and $u : X \times R \to S$
(``update''). Hence the play and coutility functions of a game
$\G : (X, S) \longrightarrow (Y, R)$ can equivalently be described as
a family of lenses
$(P_\G(\sigma, -), C_\G(\sigma, -, -)) : (X, S) \to (Y, R)$ indexed by
strategies $\sigma \in \Sigma_\G$. Further, the data involved in the
equilibrium function can be described by a ``global element'' lens
$(\One,\One) \to (X, S)$ and a ``global co-element'' lens
$(Y, R) \to (\One, \One)$. As a result, most reasoning about open
games can be done diagrammatically using that lenses also compose:
given $(v, u) : (X,S) \to (Y',R')$ and $(v', u') : (Y',R') \to (Y,R)$,
we can construct a lens $(X, S) \to (Y, R)$ by
$ (v' \circ v : X \to Y, (x, y) \mapsto u(x, u'(v(x), y)) : X \times Y
\to R) $.

There is an identity-on-objects functor
$\embedDiSet{-,-} : \Set \times \Set^{\mathrm{op}} \to \Lens$ that
maps a pair of functions $({f:X \to Y}, {g: R \to S})$ to a lens
$\embedDiSet{f,g}: (X,S) \to (Y,R)$ with $f$ as first component and
$g \circ \pi_2 : X \times R \to S$ as second component.

\begin{definition}
  Let $\G : (X,S) \longrightarrow (Y, R)$ and
  $\G' : (X',S') \longrightarrow (Y', R')$ be pure (probabilistic)
  open games. A \emph{morphism of pure (probabilistic) games}
  $\G \to \G'$ consists of functions
  \[
    (f_P : X \to X', f_C : S' \to S)
    \qquad
    (g_P : Y \to Y', g_C : R' \to R)
  \]
  and $h : \Sigma_G \to \Sigma_{G'}$, such that the following diagram
  of lenses commutes for each 
  $\sigma \in \Sigma_{G}$
  \[
    \xymatrix{
      (X, S) \ar[r]^-{\embedDiSet{f_P, f_C}} \ar[d]_-{(P_\G(\sigma), C_\G(\sigma))} & (X', S') \ar[d]^-{(P_{\G'}(h(\sigma)), C_{\G'}(h(\sigma)))} \\
      (Y, R) \ar[r]_-{\embedDiSet{g_P, g_C}} & (Y', R')
    }
  \]
  and, for every $x \in X$ and $k : Y' \to R'$, we have
  that $\sigma \in \equilib_{\G}\,x\,(g_C \circ k \circ g_p)$ implies
  \begin{itemize}
  \item $h(\sigma) \in \equilib_{\G'}\,(f_P(x))\,k$ for pure games,
  \item $\D(h)(\sigma) \in \equilib_{\G'}\,(f_P(x))\,k$ for
    probabilistic games.
  \end{itemize}
  We write $\GameCat$ and $\GameCatPure$ for the categories of
  probabilistic and pure open games, respectively, where the morphisms
  are defined as above.
\end{definition}

This is a generalisation of the definition of morphism of state-free
games used in our paper on iterated open games~\cite{iteratedGames},
but different from the notion of morphism employed by
Hedges~\cite{julesMorphisms}, which fails to make the determinisation
operation $\hatSymbol$ from \Cref{def:determ-game} a functor. As there
are currently a number of viable notions of morphisms of games (even
of lenses), we consider this empirical evidence important for what an
appropriate notion of morphism for games ought to be.  For the rest of
this section, let $\GameCatPure'$ be the category
$\GameCatPure$, except that utility and coutility sets are additionally endowed with $\D$-algebra structure.

\begin{prop}
  \label{thm:delta-functor}
  A variant of determinisation $\hatSymbol'$
  mapping a probabilistic game $\G : (X, S) \longrightarrow (Y,R)$ to a pure
  game $\hatSymbol'(\G) : (\D{}X,\D{}S) \longrightarrow (\D{}Y,\D{}R)$ (using the double strength $\laxell$, and $\liftpred{-}$),
  still with strategy set
  $\Sigma_{\hatSymbol'(\G)} = \D(\Sigma_{\G})$, extends to a functor
  $\hatSymbol' : \GameCat \to \GameCatPure'$. \hfill \QEDsymbol
\end{prop}
Determinisation $\Delta$ itself is a functor if restricted to games
whose coutility preserves the $\D$-algebra structure in a certain
sense. One might hope that one of these functors might have a left or
a right adjoint, but this is too much to ask, since it would imply in
turn that $\D$ has both a left and a right adjoint. However, we show
that the canonical way to embed a pure game as a probabilistic game
has a right adjoint.

\begin{theorem}
  \label{thm:adjunction}
  Let $\canonicalEmbed : \GameCatPure' \to \GameCat$ be the functor
  that acts as the identity on the strategy set and the lens
  structure, with
  $\equilib_{\canonicalEmbed(\G)}\,x\,k = \{ \eta(\sigma)\ |\ \sigma
  \in \equilib_{\G}\,x\,k\ \}$. Then
  \[
    \UseTwocells
    \xymatrix{\GameCat & \GameCatPure' \ltwocell_{\canonicalEmbed}^{\canonicalAdjoint}{`\perp}}
  \]
  where $\canonicalAdjoint : \GameCat \to \GameCatPure'$ similarly
  acts as the identity on the strategy set and the lens structure,
  with
  $\equilib_{\canonicalAdjoint(\HH)}\,x\,k = \{ \sigma\ |\
  \eta(\sigma) \in \equilib_{\HH}\,x\,k\ \}$. \hfill \QEDsymbol
\end{theorem}

\section{Conclusions and Future Work}

We have presented a framework for compositional game theory which
encompasses also mixed strategies, and shown that it is closed under
parallel and sequential composition, and shown that it can adequately
model common games such as Matching Pennies (where mixed strategies
are crucial) and the Market Entry Game. We also defined a notion of
morphism between games, and showed that it gives rise to a category of
games that we that can be useful for reasoning, e.g.\ by employing
adjunctions between pure and probabilistic games.

Several challenges remain. While we have accurately captured mixed
strategy Nash equilibria --- a fundamental solution concept in game
theory --- it remains to be seen if this framework can exploit the
non-independent distributions that arise naturally in it to capture
also \emph{correlated equilibria} or perhaps even \emph{evolutionary
  stable strategies}. Finally, we remark that most of our proofs do
not use any particular properties of the commutative monad $\D$. We
think this can be used to uniformly model other ``effectful''
game-theoretic phenomena such as e.g.\ quitting games using the
exceptions monad.

\paragraph{Acknowledgements}
  We thank the participants at the Third Symposium on Compositional
  Structures (SYCO3) in Oxford, March 2019 for discussions and comments.

\bibliographystyle{eptcs}
\bibliography{openGames}

\end{document}

%% file: macros.tex
\newcommand{\One}{\mathbf{1}}

\newcommand{\Pow}{\mathscr{P}}

\newcommand{\R}{\mathbb{R}}

\DeclareMathOperator*{\argmax}{arg\,max}

\newcommand{\Set}{\mathsf{Set}}

\newcommand{\Pure}{\mathsf{G}_{\mathsf{Pure}}}
\newcommand{\Prob}{\mathsf{G}_{\mathsf{Prob}}}

\newcommand{\GameCat}{\mathsf{Game}_{\mathsf{Prob}}}
\newcommand{\GameCatPure}{\mathsf{Game}_{\mathsf{Pure}}}

\newcommand{\canonicalEmbed}{\Theta}
\newcommand{\canonicalAdjoint}{\Psi}

\newcommand{\D}{\mathcal{D}}
\newcommand{\Expect}{\mathbb{E}}

\newcommand{\laxell}{\ell}

\DeclareMathOperator*{\supp}{supp}

\newcommand{\equilib}{E}

\newcommand{\G}{\mathcal{G}}
\newcommand{\HH}{\mathcal{H}}

\newcommand{\I}{\mathcal{I}}
\newcommand{\Id}{\mathcal{ID}}

\newcommand{\hatSymbol}{\Delta}
\newcommand{\hatgame}[1]{\hatSymbol(#1)}

\newcommand{\seqSymbol}{\fatsemi}
\newcommand{\seq}[2]{#1\seqSymbol #2}

\newcommand{\liftpred}[1]{\overline{\D}^{\hspace{-2pt}\scalebox{0.935}{\texttt{\#}}}\!(#1)}
\newcommand{\KleisliLift}[1]{#1^{\texttt{\#}}}

\newcommand{\Lens}{\mathsf{Lens}}

\newcommand{\embedDiSet}[1]{\iota(#1)} 

\newcommand{\e}{\sigma_{E}}
\newcommand{\de}{\sigma_{NE}}
\newcommand{\swap}{\mathsf{swap}}

\newcommand{\eval}{\mathsf{eval}}

\newtheorem{theorem}{Theorem}
\newtheorem{lemma}[theorem]{Lemma}
\newtheorem{prop}[theorem]{Proposition}
\newtheorem{fact}[theorem]{Fact}

\theoremstyle{plain}
\theorembodyfont{\upshape}
\theoremseparator{.}
\theoremsymbol{\ensuremath{\blacklozenge}}
\newtheorem{definition}[theorem]{Definition}

\theoremsymbol{\ensuremath{\color{black}\bullet\mathllap{\circ}}}
\newtheorem{example}[theorem]{Example}

\theoremheaderfont{\sc\bf}
\theorembodyfont{\upshape}
\theoremstyle{nonumberplain}
\theoremseparator{}
\newcommand{\QEDsymbol}{\rule{1ex}{1ex}}
\theoremsymbol{\QEDsymbol}
